\documentclass[useAMS,usenatbib]{mn2e}
\usepackage{graphicx}
\title[Accretion in V Puppis]{Constraints on black hole accretion in V Puppis}
\author[T.J Maccarone et al.]{Thomas J. Maccarone, Robert P. Fender, Christian Knigge\\ 
School of Physics and Astronomy, University of Southampton, Hampshire SO17 1BJ,United Kingdom\\
\newauthor
A.K. Tzioumis\\
Australia National Telescope Facility, CSIRO, PO Box 76, Epping, New South Wales 17190, Australia}

\begin{document}

\date{}

\pagerange{\pageref{firstpage}--\pageref{lastpage}} \pubyear{}

\maketitle

\label{firstpage}

\begin{abstract}
In light of the recent suggestion that the nearby
eclipsing binary star system V Puppis has a dark companion on a long
orbit, we present the results of radio and X-ray observations of it.
We find an upper limit on its radio flux of about 300 $\mu$Jy and a
detection of it in the X-rays with a luminosity of about
3$\times10^{31}$ erg/sec, a value much lower than what had been
observed in some of the low angular resolution surveys of the past.
These data are in good agreement with the idea that the X-ray emission
from V Puppis comes from mass transfer between the two B stars in the
system, but can still accommodate the idea that the X-ray emission
comes from the black hole accreting stellar wind from one or both of
the B stars.
\end{abstract}


\section{Introduction}
Nearly every black hole candidate in the Universe is either a
supermassive black hole at the center of a galaxy, or a stellar mass
black hole in an X-ray binary.  Only a few exceptions are known --
objects detected through microlensing surveys (Bennett et al 2002;
Poindexter et al. 2005).  Stellar evolution predictions indicate that
there are likely to be about $10^8$ black holes in the Galaxy of which
only about $10^3-10^4$ ever end up as low mass black hole X-ray
binaries (Romani 1992; Portegies Zwart et al. 1997).  Theoretical
modeling also suggests that binary stellar evolution will produce a
different distribution of black hole masses than will the evolution of
single stars (e.g. Fryer \& Kalogera 1998).  Thus, our knowledge of
black holes is based on a population that is both a small fraction of
the total Galactic population, and which is unrepresentative of the
larger class.

Microlensing remains one of the few means of detecting isolated black
holes (although radio emission may become useful in the LOFAR era --
Maccarone 2005).  However, black holes in extremely wide binaries may
provide a useful set of objects from which we can learn about the
properties of isolated black holes.  In any event, they are likely to
represent a substantial fraction of the black holes in the Galaxy,
since nearly all stars massive enough to form black holes are members
of multiple star systems.  In a sufficiently wide binary star system,
the stars evolved essentially independently, so the most serious
effects of multiple membership are avoided.

The first serious attempt to find observational evidence for black
holes in the Galaxy was an attempt to look for their gravitational
signatures on the orbits of other stars, rather than for evidence of
accretion (Trimble \& Thorne 1969).  However, this attempt was
unsuccessful, and in the years since, no serious attempts, in the
knowledge of the authors, have been made to search for black holes in
binary systems, apart from through searches for accretion signatures.

Recently, studies of the close eclipsing binary V Puppis have shown
periodic residuals in the eclipse timing suggesting the presence of a
$\sim$ 10 $M_\odot$ companion to the binary, with an orbital period of
about 5 years (Qian, Liao \& Fernandez-Lujas 2008).  If such an object
were anything other than a black hole, it would be observable in the
optical spectra made of V Puppis, as a third component of brightness
similar to the two known components.  Qian et al. (2008) pointed out
that Uhuru, Copernicus, and ROSAT had all seen X-ray emission from V
Puppis, at a level consistent with the level expected from accretion
of the stellar winds from V Puppis; V Puppis may represent the first
black hole in the Galaxy whose mass can be measured accurately through
Keplerian motions, but whose formation is not subject to binary
evolutionary considerations (Qian et al. 2008).  Triple systems with
accreting components have been invoked in the past -- to provide an
alternative to a black hole in Cygnus X-1 (Bahcall et al. 1975), to
explain superorbital periods in systems with accreting neutron stars
and black holes (Priedhorsky et al. 1983; Gies \& Bolton 1984;
Zdziarski et al. 2007), and to explain the presence of an F star
counterpart to 4U 2129+47, whose neutron star has an eclipse period of
5.24 hours (Garcia et al. 1989; Nowak, Heinz \& Begelman 2002).  While
the evidence for X-ray triple systems above is, in some cases, strong,
in no case is it yet definitive.

Motivated by this possibility, we performed two observations aimed at
testing the hypothesis that the X-ray emission from this system is
from accretion onto the putative black hole: searching for radio
emission from the system, since a high ratio of radio to X-ray flux is
a characteristic signature of faint accreting stellar mass black
holes; and making an X-ray image of this field with Chandra, to ensure
that the X-rays really are coming from V Puppis and not some other
nearby object.

\section{Data used, analysis procedure, and results}
\subsection{Australia Telescope Compact Array data}

The Australia Telescope Compact Array observed V Puppis for 2.97 hours
on the date of 17 July 2008, simultaneously at 4.80 and 8.64 GHz,
using B1934-638 and B0823-500 as flux and phase calibrators,
respectively.  No emission from V Puppis was detected.  The rms noise
levels were 0.1 mJy at 4800 MHz, and 0.08 mJy at 8640 MHz.

\subsection{Chandra data}
We observed V Puppis with the Chandra X-ray Observatory for 1170
seconds on 1 September 2008.  The High Resolution Camera (HRC) was
used, because its detector is a microchannel plate, rather than a
charged coupled device (CCD).  Due to optical loading on CCDs, it is
not possible to use CCDs to observe an X-ray source which is optically
as bright as V Puppis without serious problems with the resultant data
quality.  We detected 73 photons from within 5 arcseconds on V Puppis.
The background level in this area in this exposure with the HRC is
about 4 photons.  An X-ray image in a region around V Puppis is shown
in figure 1.  Using HEASARC's WebPIMMS, and assuming a Galactic
foreground $N_H$ of $1.5\times10^{21}$ cm$^{-2}$ and a power law
spectrum with photon index of 1.7, we make an estimate of the flux of
V Puppis, and find an unabsorbed X-ray flux of $2\times10^{-12}$ erg
s$^{-1}$ cm$^{-2}$, giving an X-ray luminosity at 300 pc of $2\times
10^{31}$ erg s$^{-1}$.  Unfortunately, it is not possible to obtain
reliable spectral information from the Chandra HRC, and no previous
observations with reliable spectroscopy had sufficient image quality
to resolve apart V Puppis and the other source visible in the figure,
which is about 20'' away, so the uncertainty in the luminosity is
dominated by the uncertainty about which spectral model should be
assumed.

\begin{figure}
\includegraphics[height=4 cm]{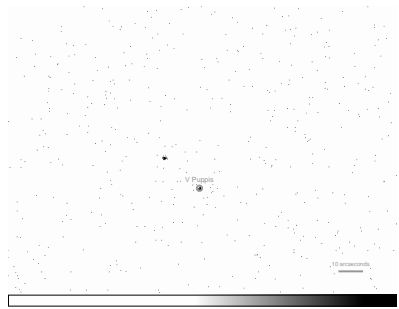}
\caption{The X-ray image from the Chandra HRC of the region around V
Puppis.  The circle is about 1 arcsecond in size, centered on the
position of V Puppis from SIMBAD.  A line segment of 10 arcseconds is
shown in the lower left of the figure for scale.}
\end{figure}

\section{Discussion}

The X-ray luminosity of the V Puppis system, as inferred from these
data, is similar to the X-ray luminosities of Algol type systems, and
quite a bit fainter than the faintest colliding wind binaries.
Following Frank, King \& Raine (1995), we can estimate that for a 10
$M_\odot$ black hole in a 5.5 year orbit around a binary of about 25
$M_\odot$ in total mass, the X-ray luminosity should be $\sim10^{31}$
erg/sec for a mass loss rate of $\sim10^{-5}M_\odot$ yr$^{-1}$,
albeit with considerable uncertainty in several key parameters making
this estimate uncertain by multiple orders of magnitude. Both Uhuru
and Copernicus observed much higher X-ray luminosities from V Puppis
than what we observe here, while our observations give a similar
luminosity to that observed with ROSAT.  The Uhuru measurement gives a
count rate of $9.4\pm2.3$ counts s$^{-1}$, equivalent to an X-ray flux
of 1.6$\times10^{-10}$ erg s$^{-1}$ cm$^{-2}$, assuming a standard
spectral shape (Giacconi et al. 1974), yielding an X-ray luminosity of
$2\times10^{33}$ erg s$^{-1}$, at the distance of V Puppis.
Copernicus observed a similar X-ray luminosity from the same region on
the sky, but given that one cannot be sure whether the emission came
from V Puppis or a nearby object, one cannot make as accurate an
estimate of the luminosity from Copernicus, since one needs to know
the source position to good accuracy in order to estimate the
collimator response (Bahcall et al. 1975).  This may be indicative of
variability of the X-ray luminosity of either V Puppis or of one of
the other sources within the roughly square degree error boxes of V
Puppis as seen with Uhuru and Copernicus.  If the putative black hole
in V Puppis is in a sufficiently eccentric orbit, its Bondi luminosity
could change by orders of magnitude over its orbit.  The detection in
the ROSAT all sky survey (Voges et al. 1999) is at a level which is
about twice as bright as what we see with Chandra, under the same
assumptions.  There is another X-ray source, at the position of the
star HJ 4025C (about 20'' from V Puppis), which is about 75\% as
bright in X-rays as is V Puppis and within the ROSAT PSPC error box;
there is little evidence for substantial variability between the ROSAT
and Chandra observations.

The combination of the X-ray and radio emission argues that not all of
the X-ray emission is likely to be coming from a 10 $M_\odot$ black
hole.  Using the most recent formulation (K\"oerding, Falcke \& Corbel
2005) of the fundamental plane of black hole activity
(Merloni, Heinz \& Di Matteo 2003; Falcke, K\"ording \& Markoff 2004),
we find that a 10 $M_\odot$ black hole with an X-ray flux of
$2\times10^{-12}$ erg/sec/cm$^2$ at a distance of 300 pc should have
a radio flux of about 4 mJy.  The X-ray/radio correlation of black
holes only (Gallo, Fender \& Pooley 2003) predicts a flux of about 2
mJy.  The rms scatter in K\"ording et al (2005) of about a factor of 3
is still not large enough to explain the difference between the
observed radio flux and the radio flux predicted based on the X-ray
luminosity.  Therefore, at least a substantial fraction of the X-ray
emission we observed was most likely from the mass transfer between
the two B star components of V Puppis.  However, only deeper radio
observations can rule out the possibility that some reasonably large
fraction of the emission is from accretion onto the candidate black
hole.

\section*{Acknowledgments}
We thank Harvey Tananbaum for granting our Chandra Director's
Discretionary Time request.  The Australia Telescope is funded by the
Commonwealth of Australia for operation as a National facility managed
by CSIRO.

\label{lastpage}

\end{document}